\documentclass[twocolumn,pra,amsmath,amssymb]{revtex4-1}
\usepackage[latin9]{inputenc}
\usepackage{color}
\usepackage{amsmath}
\usepackage{amssymb}
\usepackage{graphicx}
\usepackage[unicode=true,pdfusetitle,
 bookmarks=true,bookmarksnumbered=false,bookmarksopen=false,
 breaklinks=false,pdfborder={0 0 1},backref=false,colorlinks=false]
 {hyperref}
\begin{document}
\title{Emergent criticality and universality class of the finite temperature
charge density wave transition in lattice Bose gases within optical
cavities}
\author{Liang He}
\email{liang.he@scnu.edu.cn}

\affiliation{Guangdong Provincial Key Laboratory of Quantum Engineering and Quantum
Materials, SPTE, South China Normal University, Guangzhou 510006,
China}
\author{Su Yi}
\email{syi@itp.ac.cn}

\affiliation{CAS Key Laboratory of Theoretical Physics, Institute of Theoretical
Physics, Chinese Academy of Sciences, Beijing 100190, China}
\affiliation{School of Physical Sciences \& CAS Center for Excellence in Topological
Quantum Computation, University of Chinese Academy of Sciences, Beijing
100049, China}
\begin{abstract}
We investigate the finite temperature charge density wave (CDW) transition
of lattice Bose gases within optical cavities in the deep Mott-insulator
limit. We find a new critical regime emerges at a temperature around
one-half of the on-site interaction energy, where the first-order
CDW transition at low temperatures terminates at a critical point
and changes to a second-order one. By directly calculating the critical
exponents and constructing the effective theory in the corresponding
critical regime, we find the emergent criticality belongs to the five-dimensional
Ising universality class. Direct experimental observation of the emergent
criticality can be readily performed by current experimental set-ups
operated in the temperature regime around half the on-site interaction
energy. 
\end{abstract}
\maketitle

\section{Introduction}

Long-range interactions can give rise to rich exotic structures and
phases of matter, such as charge and spin density waves, supersolids,
spin-glasses, etc. Moreover, on the fundamental level, the long-range
characteristic of interactions can play the same crucial role just
as symmetries and spatial dimensions of physical systems in determining
their universal physical behavior in the critical regime of their
continuous phase transitions \citep{Fisher_RMP_1974,Wilson_RMP_1975}.
In the context of ultracold atoms, various long-range interacting
systems, ranging from ultracold gases with large magnetic or electric
dipole moments \citep{Stuhler_PRL_2005,Ni_Science_2008}, over atoms
in Rydberg states \citep{Heidemann_PRL_2008}, to ultracold gases
in cavities with cavity-photon-mediated interactions \citep{Baumann_nature_2010,Mottl_Science_2012,Landig_Nature_2016},
have been realized in experiments \citep{Stuhler_PRL_2005,Heidemann_PRL_2008,Ni_Science_2008,Baumann_nature_2010,Mottl_Science_2012,Landig_Nature_2016},
making them powerful platforms to explore the fundamental behavior
characteristic of long-range interactions.

A case in point is Bose gases in two-dimensional (2D) square optical
lattices within optical cavities, which feature distinct \emph{infinite}-long-range
(ILR) interactions that are mediated by the cavity photons \citep{Landig_Nature_2016}.
Recent experimental and theoretical investigations \citep{Landig_Nature_2016,Ritsch_RMP_2013,Li_PRA_2013,Liao_PRA_2018,Niederle_PRA_2016,Panas_PRB_2017,Sundar_PRA_2016,Chen_PRA_2016,Dogra_PRA_2016,Habibian_PRL_2013}
have shown that at low temperatures this system can support rich phases
and phase transitions attributed to its long-range interaction, such
as supersolids, charge density waves (CDW), etc. In particular, by
tuning the relative strength between the short-range on-site interaction
and the ILR one in the deep Mott-insulator regime, a new phase transition
characteristic of a first-order one between the $\mathbb{Z}_{2}$-symmetric
homogeneous Mott-insulator and the spontaneous $\mathbb{Z}_{2}$-symmetry
breaking CDW phase was observed in experiments \citep{Landig_Nature_2016}.
Noticing that current experiments are mostly operated at a temperature
scale that is much lower than all other energy scales in the system,
it is intriguing to expect that at the evenly matched temperature
scale, due to the interplay among short-range on-site interactions,
ILR interactions, and thermal fluctuations, a completely different
scenario for the CDW transition could arise. This thus raises the
fundamental question of whether criticalities for the CDW transition,
whose existence is excluded at low temperatures in the first-order
transition scenario, could emerge and bear the characteristic of the
ILR interaction.

\begin{figure}
\includegraphics[width=3.36in]{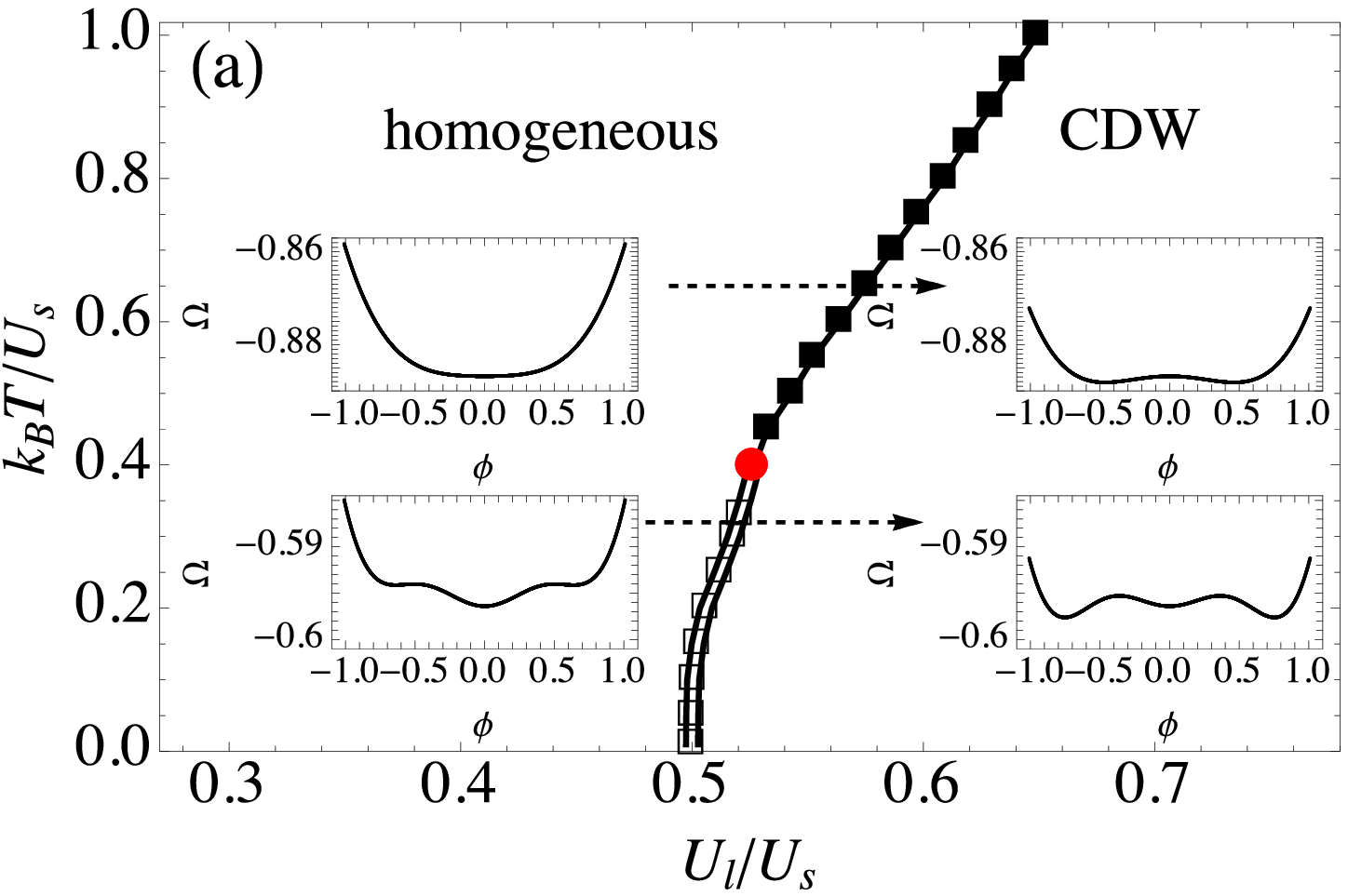}

\includegraphics[width=1.68in]{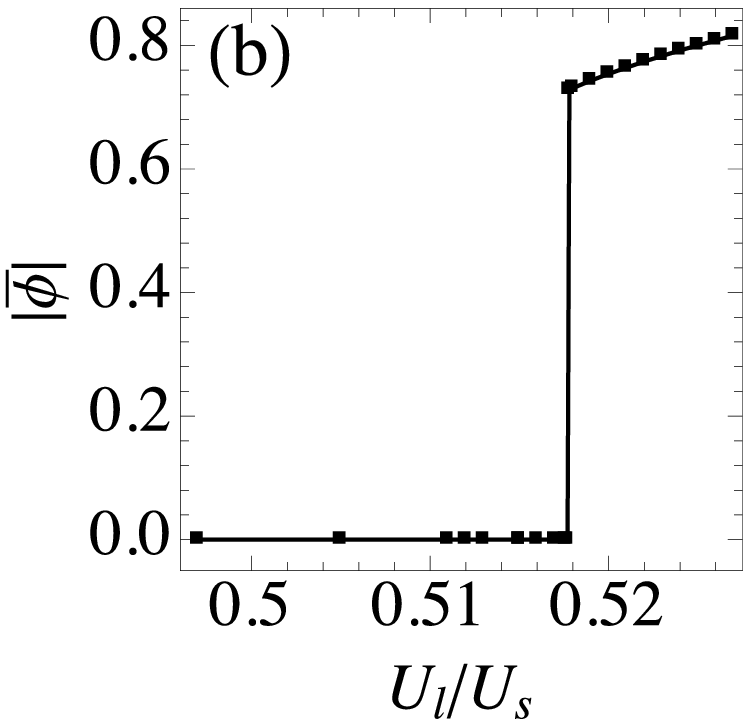}\includegraphics[width=1.68in]{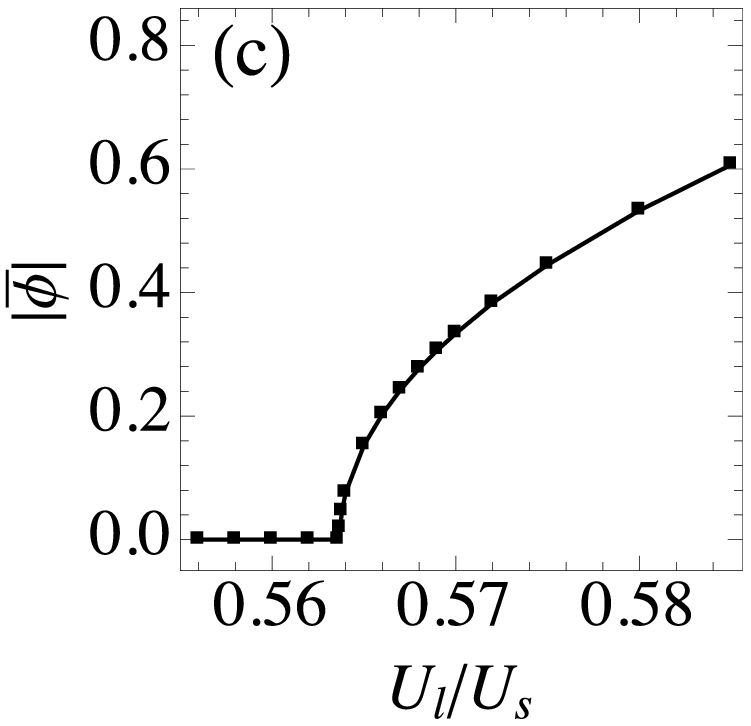}

\caption{(a) Finite temperature phase diagram of lattice Bose gases within
optical cavities in the deep Mott-insulator limit at unit filling.
In the weak ILR interaction strength $U_{l}$ regime the system is
in the homogeneous phase, while in the strong ILR interaction regime,
the system is in the charge density wave (CDW) phase, where the system's
density distribution assumes the checker-board pattern. At low temperatures,
the CDW transition between the homogeneous phase and the CDW phase
upon tuning ILR interaction strength $U_{l}$ is a first-order phase
transition (marked by open squares and a double solid curve). At high
temperatures, the first-order transition boundary terminates at a
critical point (marked by the filled red disk) whose temperature $T_{\mathrm{CP}}=0.396$
(in the unit of $U_{s}/k_{B}$, with an error less than $10^{-3}$).
Above the critical point, i.e., $T>T_{\mathrm{CP}}$, the CDW transition
becomes a second-order one (marked by solid squares and a solid curve).
The insets in the phase diagram show the dependence of $\Omega$ on
the CDW order parameter field $\phi$ in four typical scenarios, namely,
the homogeneous phase above (below) $T_{\mathrm{CP}}$ {[}the upper-left
(lower-left) inset{]} and the CDW phase above (below) $T_{\mathrm{CP}}$
{[}the upper-right (lower-right) inset{]}. (b) The ILR interaction
strength $U_{l}$ dependence of the CDW order parameter $|\bar{\phi}|$
{[}lower dashed arrow in (a){]} at a temperature below $T_{\mathrm{CP}}$
with $k_{B}T=0.3U_{s}$, showing the first-order transition upon increasing
$U_{l}$. (c) The ILR interaction strength $U_{l}$ dependence of
the CDW order parameter $|\bar{\phi}|$ {[}upper dashed arrow in (a){]}
at a temperature above $T_{\mathrm{CP}}$ with $k_{B}T=0.6U_{s}$,
showing the second-order transition upon increasing $U_{l}$. See
text for more details.}
\label{Fig_grand_canonical_Phase_diagram} 
\end{figure}

In this paper, we address the above question by establishing the complete
finite-temperature phase diagram of the system in the deep Mott-insulator
limit at unit filling {[}cf.~Fig.~\ref{Fig_grand_canonical_Phase_diagram}(a){]}
and investigating the emergent critical scaling behavior of the system
{[}cf.~Fig.~\ref{Fig_tricritical_point} and Fig.~\ref{Fig_critical_exponent_2nd_order_transition}{]}.
More specifically, we find the following. (i) An emergent critical
regime that consists of a new critical point and second-order CDW
transitions. At low temperatures, our calculations clearly show that
the CDW transition is a first-order phase transition, i.e., the CDW
order parameter $\bar{\phi}$ assumes a finite jump $\Delta\bar{\phi}$
when the transition boundary is crossed {[}cf.~Figs.~\ref{Fig_grand_canonical_Phase_diagram}(a)
and \ref{Fig_grand_canonical_Phase_diagram}(b){]}, which corroborates
observations in experiments \citep{Landig_Nature_2016}. When the
temperature is increased, the jump of the CDW order parameter $\Delta\bar{\phi}$
decreases and finally vanishes at a critical point with its temperature
$T_{\mathrm{CP}}=0.39U_{s}/k_{B}$ {[}$U_{s}$ and $k_{B}$ are the
on-site energy strength and the Boltzmann constant, respectively.
cf.~Fig.~\ref{Fig_grand_canonical_Phase_diagram}(a) and Fig.~\ref{Fig_tricritical_point}{]}.
Above the critical point, the CDW transition becomes a second-order
transition, where the CDW order parameter changes continuously when
crossing the transition boundary {[}cf.~Figs.~\ref{Fig_grand_canonical_Phase_diagram}(a)
and \ref{Fig_grand_canonical_Phase_diagram}(c){]}. (ii) The universality
class of the emergent criticality belongs to the five-dimensional
(5D) Ising universality class. The CDW order parameter jump $\Delta\bar{\phi}$
along the first-order CDW transition boundary assumes a power law
scaling with respect to the temperature change near the critical point,
i.e., $\Delta\bar{\phi}\propto(T_{\mathrm{CP}}-T)^{0.5}$ (cf.~Fig.~\ref{Fig_tricritical_point}).
Moreover, the CDW order parameter also shows the same power law scaling
near the second-order CDW transition boundary, i.e., $|\bar{\phi}|\propto(T_{c}-T)^{0.5}$
with $T_{c}$ being the critical temperature at the second-order transition
boundary (cf.~Fig.~\ref{Fig_critical_exponent_2nd_order_transition}).
Analyses of the effective theory in the critical regime {[}cf.~Eqs.~(\ref{eq:GL_free_energy},~\ref{eq:Exponents_from_GL}){]}
show these critical scaling behavior of this low dimensional 2D system
belong to the universality class of short-range interacting systems
with a much higher spatial dimension, i.e., 5D Ising universality
class. This clearly shows that the criticality of the system is strongly
influenced by and thus bear the long-range characteristic of its interactions.
Moreover, as far as we know, this also establishes lattice Bose gases
in optical cavities as the first type of realistic physical systems
that accommodates exact physical manifestations of the largely academic
5D Ising universality class \citep{sidenote1}.

\section{Model in the deep Mott-insulator limit}

For Bose gases in optical lattices located inside optical cavities,
besides the conventional on-site interaction, the strong coupling
between cavity photons and bosonic atoms can result in an effective
ILR interaction for Bose gases \citep{Mottl_Science_2012,Landig_Nature_2016}.
Their physics in a wide range of the parameter space can be captured
by the ILR interacting Bose-Hubbard model (cf.~Ref.~\citep{Landig_Nature_2016}
for detailed derivations), whose Hamiltonian consists of a conventional
hopping part and an interaction part. In this work, we focus on the
physics in the deep Mott-insulator limit, where the hopping amplitude
is negligibly small, hence the system is described by the interaction
part alone. In this limit, its Hamiltonian reads 
\begin{equation}
\hat{H}=\frac{U_{s}}{2}\sum_{i,\sigma}\hat{n}_{i,\sigma}(\hat{n}_{i,\sigma}-1)-\frac{U_{l}}{L}\left(\sum_{i=1}^{L/2}\hat{n}_{i,e}-\sum_{i=1}^{L/2}\hat{n}_{i,o}\right)^{2}.\label{eq:Hamiltonian}
\end{equation}
Here, the first term describes the conventional onsite interaction
with its strength characterized by $U_{s}$. The second term describes
the ILR interaction mediated by photons in the cavity \citep{Mottl_Science_2012,Landig_Nature_2016}
with its strength characterized by $U_{l}$. Moreover, in order to
restore the conventional thermodynamical limit, $U_{l}$ is further
rescaled by the total number of lattice sites $L$ in this term according
to the Kac prescription \citep{Kac_J_Math_Phys_1963}. Here, we consider
the 2D square lattice case which is the same as the experimental set-up
in Ref.~\citep{Landig_Nature_2016}, and refer to its two interpenetrating
square sub-lattices as ``even'' ($e$) and ``odd'' ($o$) lattice,
respectively. $\hat{n}_{i,\sigma}$ is the particle number operator
that counts the number of atoms at site $i$ on the sub-lattice $\sigma$,
with $\sigma=e,\,o$. We remark here although to be concrete, we base
our discussion on the 2D square lattice which is most relevant for
the current experimental set-ups \citep{Landig_Nature_2016}, the
results to be presented in the following generally hold true for generic
bipartite lattices.

From the Hamiltonian (\ref{eq:Hamiltonian}), we see that at fixed
integer filling, the short-range on-site interaction term favors a
conventional Mott-insulator phase where the particle density is homogeneously
distributed over the lattice, while the ILR interaction term favors
a CDW phase with a chequerboard pattern where particle densities on
the ``even'' and ``odd'' chequerboard sub-lattice are different,
thus breaking the $\mathbb{Z}_{2}$-symmetry between the two sub-lattices
\citep{Mottl_Science_2012}. The competition between these two types
of interactions, hence two energy scales, gives rise to a phase transition
associated with the $\mathbb{Z}_{2}$-symmetry breaking in the deep
Mott-insulator limit as observed in experiments focusing at fixed
low temperatures \citep{Landig_Nature_2016}. Taking into account
the energy scale set by the temperature, one would expect the competition
among these three energy scales could give rise to new physics beyond
the one in the low-temperature regime. Indeed, as we shall see in
the following, when the energy scale associated with the temperature
can match the two other energy scales in the system, new\emph{ }criticalities
that are dominated by the ILR interaction emerge.

\section{Emergent criticality for the CDW transition at intermediate temperature
scales}

Before discussing the main results, let us briefly outline the major
method we used in our calculations. To investigate the finite temperature
phase transition between the CDW and the homogeneous phase (at low
temperatures this corresponds to the homogeneous Mott-insulator),
we introduce the CDW order parameter field $\phi$ into the quantum
grand partition function $Z$ of the system via the standard Hubbard-Stratonovich
transformation and reformulate the grand partition function $Z$ in
terms of the $\phi$ field (cf.~Appendix~\ref{sec:Appendix_HST}),
whose explicit form reads 
\begin{equation}
Z=\sqrt{\frac{\beta U_{l}L}{\pi}}\int_{-\infty}^{+\infty}d\phi\,e^{-\beta L\Omega{}_{\{\beta,\mu,U_{s},U_{l}\}}(\phi)}\label{eq:Z_expressed_in_phi}
\end{equation}
with 
\begin{align}
 & \,\Omega{}_{\{\beta,\mu,U_{s},U_{l}\}}(\phi)\nonumber \\
\equiv & \,U_{l}\phi^{2}-\frac{1}{2\beta}\sum_{\eta=\pm1}\ln\left[\sum_{n=0}^{+\infty}e^{-\beta\left[\frac{U_{s}}{2}n(n-1)-\mu n+2\eta U_{l}n\phi\right]}\right].\label{eq:Omega_tilde}
\end{align}
Here, $\mu$ is the chemical potential and $\beta=\left(k_{B}T\right)^{-1}$
with $k_{B}$ being the Boltzmann constant and $T$ being the temperature.
The transition from the homogenous phase to the CDW is characterized
by the appearance of the non-zero expectation value of $\phi$, i.e.,
CDW order parameter $\bar{\phi}\equiv\langle\phi\rangle=\langle\sum_{i=1}^{L/2}\hat{n}_{i,e}-\sum_{i=1}^{L/2}\hat{n}_{i,o}\rangle/L$
(cf.~Appendix~\ref{sec:Appendix_HST}). In the thermodynamic limit
$L\rightarrow\infty$, the integral with respect to $\phi$ in Eq.~(\ref{eq:Z_expressed_in_phi})
is given exactly by its saddle point integration. Therefore, in the
thermodynamic limit, $Z=\left(\sqrt{\beta U_{l}L/\pi}\right)\exp\left(-\beta L\min\left[\Omega{}_{\{\beta,\mu,U_{s},U_{l}\}}(\phi)\right]\right)$
and the CDW order parameter $\bar{\phi}$ is given by the value of
$\phi$ that minimizes $\Omega{}_{\{\beta,\mu,U_{s},U_{l}\}}(\phi)$.
The summation in Eq.~(\ref{eq:Omega_tilde}) can not be performed
analytically, however, it can be numerically calculated at a sufficiently
high accuracy with a large enough cut-off on $n$. This enables us
to map out the complete finite temperature phase diagram as we shall
now discuss.

At unit filling, the finite temperature phase diagram is shown in
Fig.~\ref{Fig_grand_canonical_Phase_diagram}(a). In the low-temperature
regime (compared to half of the on-site energy), the transition from
the homogenous phase to the CDW phase is a first-order transition,
where the CDW order parameter shows a finite jump $\Delta\bar{\phi}$
when the system parameter is tuned across the transition boundary
{[}cf.~Fig.~\ref{Fig_grand_canonical_Phase_diagram}(b){]}. This
corroborates the findings in experiments where hysteretic behavior
of the first-order CDW transition was observed in the low-temperature
regime \citep{Landig_Nature_2016}. The first-order transition behavior
can be traced back to the structure of the function $\Omega{}_{\{\beta,\mu,U_{s},U_{l}\}}(\phi)$:
For the homogeneous phase, $\Omega{}_{\{\beta,\mu,U_{s},U_{l}\}}(\phi)$
has two types of minimums, with one type minimum located at $\phi=0$
which is global and the other type located at $\pm\phi^{*}$, with
$|\phi^{*}|\neq0$ which is local {[}cf. lower-left inset in Fig.~\ref{Fig_grand_canonical_Phase_diagram}(a){]}.
When system parameters are tuned to approach the first-order transition
boundary the difference in the $\Omega$ value between these two types
of minimums decreases. At the first-order transition boundary, $\Omega$
assumes the same value at these two types of minimums. After the system
parameter enters the CDW regime, minimums at $\pm\phi^{*}$ become
the global minimums {[}cf. lower-right inset in Fig.~\ref{Fig_grand_canonical_Phase_diagram}(a){]},
giving rise to the finite jump in the CDW order as shown in Fig.~\ref{Fig_grand_canonical_Phase_diagram}(b).

When the temperature is increased in the low-temperature regime, the
order parameter jump $\Delta\bar{\phi}$ at the first-order transition
boundary decreases and finally vanishes at a critical point as shown
by the red dot in Fig.~\ref{Fig_grand_canonical_Phase_diagram}(a)
with its temperature $T_{\mathrm{CP}}=0.396U_{s}/k_{B}$. The emergence
of the critical point can be traced back to the change of locations
of the minimums of $\Omega$ along the first-order transition boundary
as shown in the inset of Fig.~\ref{Fig_tricritical_point}(a), where
$|\phi^{*}|$ approaches $0$ when the temperature is increased.

Above the critical point, the CDW transition becomes a second-order
phase transition, where the CDW order parameter changes continuously
when system parameters are tuned across the transition boundary as
shown in Fig.~\ref{Fig_grand_canonical_Phase_diagram}(c). The second-order
transition behavior can also be traced back to the structure of the
function $\Omega{}_{\{\beta,\mu,U_{s},U_{l}\}}(\phi)$: On the homogeneous
phase side of the transition, $\Omega{}_{\{\beta,\mu,U_{s},U_{l}\}}(\phi)$
has only one global minimum located at $\phi=0$ {[}cf. upper-left
inset in Fig.~\ref{Fig_grand_canonical_Phase_diagram}(a){]}. When
the system parameters are tuned across the second-order transition
boundary, this minimum continuously changes to a maximum, and two
new minimums emerging at $0^{\pm}$ at the same time. These two non-zero
minimums continuous moving far away from $\phi=0$ as system parameters
are further tuned into deeper CDW parameter regime {[}cf. upper-right
inset in Fig.~\ref{Fig_grand_canonical_Phase_diagram}(a) and Fig.~\ref{Fig_grand_canonical_Phase_diagram}(c){]}.

The emergence of the critical point and the second-order CDW transition
in fact gives rise to a new critical regime that is absent at low
temperatures where the CDW transition is a first-order transition.
Indeed, as we shall see in the following, in the vicinity of the critical
point and the second-order transition, both the order parameter jump
$\Delta\bar{\phi}$ of the \emph{first-order} transition and the order
parameter $\bar{\phi}$ manifest critical power law scaling that is
dominated by the long-range interaction of the system.

\section{Critical scaling and universality class of CDW transition at intermediate
temperatures}

\begin{figure}
\includegraphics[height=1.75in]{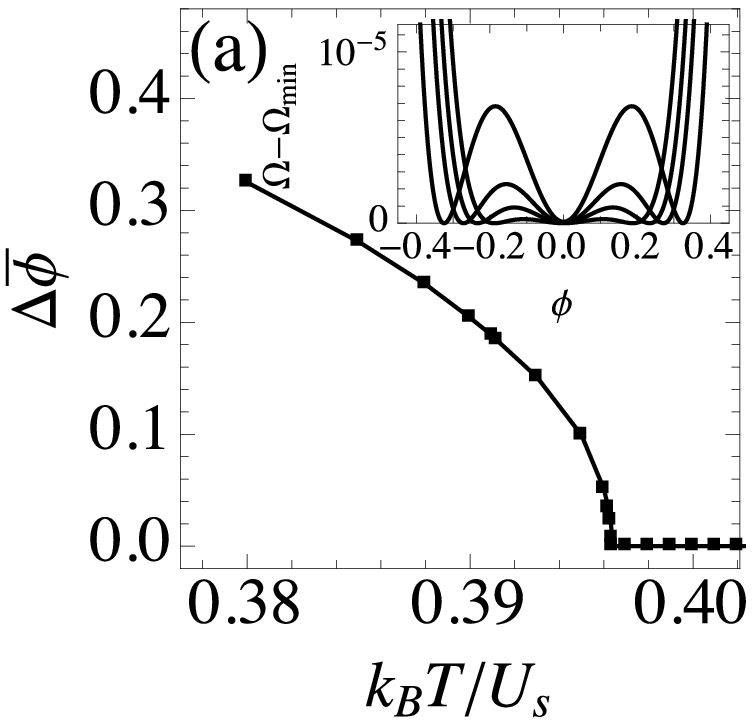}\includegraphics[height=1.75in]{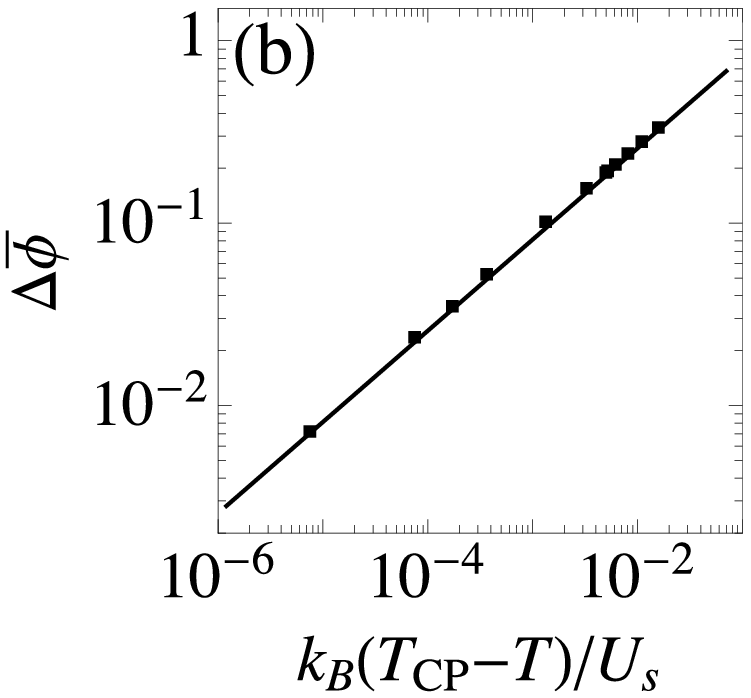}

\caption{(a) Temperature dependence of $\Delta\bar{\phi}$ along the first-order
CDW transition boundary. Upon increasing $T$ from the low temperature
regime with $T<T_{\mathrm{CP}}$, $\Delta\bar{\phi}$ continuously
decreases to zero at the temperature of the critical point $T_{\mathrm{CP}}=0.396U_{s}/k_{B}$.
The inset shows $\phi$ dependence of $\Omega$ at four different
temperatures. The curves with their double-well located from outer
position to inner position correspond to $k_{B}T/U_{s}=0.380,0.385,0.388,0.391$,
respectively. $\Omega_{\mathrm{min}}$ denotes the minimum value of
$\Omega$ for each curve. (b) Linear fit to the data points in (a)
near the critical point with $T<T_{\mathrm{CP}}$ on the double logarithmic
scale, showing clearly a power law dependence of $\Delta\bar{\phi}$
on $T_{\mathrm{CP}}-T$, i.e., $\Delta\bar{\phi}\propto(T_{\mathrm{CP}}-T)^{\alpha}$
with $\alpha$$=0.500$. The solid line correspond to the best power
law fit $\Delta\bar{\phi}\propto(T_{\mathrm{CP}}-T)^{0.500}$ to the
data points in the plot. See text for more details.}
\label{Fig_tricritical_point} 
\end{figure}

At low temperatures, the CDW order changes its value abruptly by $\Delta\bar{\phi}$
when system parameters, for instance, $U_{l}$, are tuned across the
1st order transition boundary. Therefore, the CDW order parameter
$\bar{\phi}$ itself does not show any critical power law scaling.
However, as we can see from Fig.~\ref{Fig_tricritical_point}(a),
where numerical results of the temperature dependence of the order
parameter jump $\Delta\bar{\phi}$ are shown, $\Delta\bar{\phi}$
decreases continuously upon increasing the temperature and finally
vanishes at the critical point. This thus gives rise to the possible
existence of the critical power scaling concerning the order parameter
jump $\Delta\bar{\phi}$ near the critical point. Indeed, as shown
in Fig.~\ref{Fig_tricritical_point}(b), a power law fit to the temperature
dependence of $\Delta\bar{\phi}$ in the vicinity of the critical
point $T_{\mathrm{CP}}$ clearly shows a critical scaling $\Delta\bar{\phi}\propto(T_{\mathrm{CP}}-T)^{0.500}$.

\begin{figure}
\includegraphics[width=3.3in]{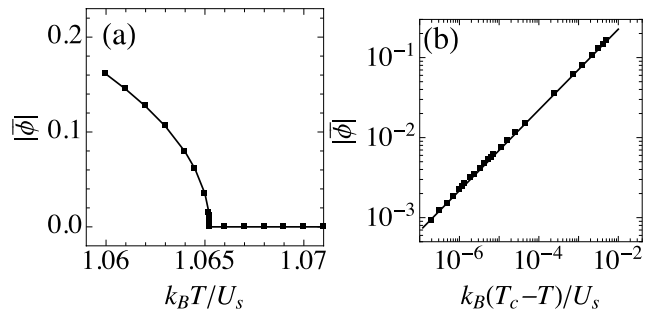}

\caption{(a) Temperature dependence of the CDW order parameter $|\bar{\phi}|$
at a fixed ILR interaction strength with $U_{l}/U_{s}=0.66$. Upon
increasing the temperature, $|\bar{\phi}|$ continuously decreases
to zero at the critical temperature $T_{c}$ ($k_{B}T_{c}/U_{s}=1.065$)
of the second-order CDW transition point. (b) Linear fit to the data
points in (a) with $T<T_{c}$ on the double logarithmic scale, showing
clearly a power law dependence of $|\bar{\phi}|$ on $T_{c}-T$, i.e.,
$|\bar{\phi}|\propto(T_{c}-T)^{\alpha}$ with $\alpha=0.504$. The
solid line corresponds to the best power law fit $|\bar{\phi}|\propto(T_{c}-T)^{0.504}$
to the data points in the plot. See text for more details.}
\label{Fig_critical_exponent_2nd_order_transition} 
\end{figure}

Above the critical point, i.e., $T>T_{\mathrm{CP}}$, the CDW transition
becomes a second-order one, where CDW order changes continuously when
system parameters are tuned across the transition boundary. Thus,
one naturally expects the CDW order $\bar{\phi}$ shows critical scaling
near the second-order transition boundary. Indeed, as we can see from
Fig.~\ref{Fig_critical_exponent_2nd_order_transition}, where numerical
results of the temperature dependence of CDW order parameter $|\bar{\phi}|$
at a fixed $U_{l}$ are shown, a critical scaling $|\bar{\phi}|\propto(T_{c}-T)^{0.504}$
can be clearly observed {[}cf.~Fig.~\ref{Fig_critical_exponent_2nd_order_transition}(b){]}.

Interestingly, the numerical values for these two critical exponents,
i.e., the one that governs the scaling of $\Delta\bar{\phi}$, which
assume the value of $0.500$, and the other one that governs the scaling
of $|\bar{\phi}|$, which assumes the value of $0.504$, respectively,
are remarkably close to each other. This thus strongly suggests these
two critical scaling behavior are related to each other on the fundamental
level by the same effective theory in their respective critical regime.
Indeed, both scaling can be determined via the same effective theory
within the Ginzburg-Landau (GL) framework, as we shall now discuss.

In the critical regime, the CDW order parameter is small enough to
allow a systematic expansion of the system's free energy $F$ with
respect to its CDW order parameter $\bar{\phi}$. The $\mathbb{Z}_{2}$
symmetry of the system determine the allowed terms in the expansion,
whose explicit form up to the sixth order in $\bar{\phi}$ reads 
\begin{equation}
F=\frac{1}{2}r\bar{\phi}^{2}+\frac{1}{4}u_{4}\bar{\phi}^{4}+\frac{1}{6}u_{6}\bar{\phi}^{6},\label{eq:GL_free_energy}
\end{equation}
with $r$, $u_{4}$, and $u_{6}$ being the GL coefficients. To describe
the first-order transition, one further assumes: $r$ depends on temperature
linearly, i.e., $r=a(T_{\mathrm{CP}}-T)$ where $a$ is a positive
coefficient, and $u_{4,6}$ are temperature independent coefficients
with $u_{4}$ assumed to be negative for $T<T_{\mathrm{CP}}$ and
$u_{6}$ being always positive in order to stabilize the whole system.
While to describe the second-order transition above the critical point,
one assumes: $r$ depends on temperature linearly, i.e., $r=a(T-T_{c})$
with $T_{c}$ being the critical temperature of the second-order CDW
transition, $u_{4}$ is assumed to be positive, hence the sixth order
term in Eq.~(\ref{eq:GL_free_energy}) is thus irrelevant in this
case. By analyzing the saddle points of $F$ in these two cases, we
can obtain: for the first-order transition, the CDW order parameter
jump $\Delta\bar{\phi}=(1/2)\sqrt{r/\left|u_{4}\right|}$ ; for the
second-order transition, $|\bar{\phi}|=\sqrt{|r|/u_{4}}$ in the ordered
phase (cf.~Appendix~\ref{sec:GL-theory}). Noticing in both cases,
$r$ is linear dependent on the temperature and $u_{4}$ is independent
of the temperature, we directly get the following scaling laws 
\begin{align}
 & \Delta\bar{\phi}\propto(T_{\mathrm{CP}}-T)^{1/2}\,\,\mathrm{and}\,\,|\bar{\phi}|\propto(T_{c}-T)^{1/2},\label{eq:Exponents_from_GL}
\end{align}
showing remarkable agreements with the numerical results on these
two scaling.

At first sight, this good agreement seems quite unexpected, since
due to the fact that long-range fluctuations are neglected in the
effective theory within the GL framework, it is only expected to provide
very rough estimations on the critical exponents for the 2D system
under consideration. However, on the other hand, noticing the ILR
interaction can strongly suppress the long-range fluctuations \citep{Bouchet_Physica_A_2010},
this in fact promotes the GL effective theory to a precise effective
theory that captures the critical scaling behavior. Such a promotion
of the same GL effective theory to a precise effective critical theory
is reminiscent of what happens in the 5D Ising model with the same
$\mathbb{Z}_{2}$-symmetry whose corresponding scaling exponent is
exactly $1/2$ \citep{Aizenman_PRL_1981} as in Eq.~(\ref{eq:Exponents_from_GL}).
In the case of the 5D Ising model, the promotion is accomplished via
suppressing the long-range fluctuations by the higher dimensionality,
while in contrast, it is the long-range interaction that suppresses
the long-range fluctuations in lattice bose gases in cavities. This
also concludes its emergent criticality belongs to the 5D Ising universality
class, which clearly shows that the criticality of the CDW transition
is strongly influenced by the long-range characteristic of the interaction
in the system.

\section{Conclusions}

The CDW transition of lattice Bose gases in optical cavities is crucially
influenced by the thermal fluctuations above the temperature around
half the on-site interaction energy: the first-order CDW transition
at low temperatures terminates at a critical point where it changes
to a second-order phase transition. This gives rise to the new emergent
criticality belonging to the 5D Ising universality class, manifesting
clearly the long-range characteristic of the system's interaction.
Noticing the CDW order parameter can be well measured in current experiments
\citep{Landig_Nature_2016}, we expect the physics in the emergent
critical regime predicted in this work can be readily observed by
operating current experimental set-ups at a temperature scale around
one-half of the on-site energy, or alternatively, by lowering both
$U_{s}$ and $U_{l}$ in experiments to effectively increase the temperature.
Moreover, noticing that even the measurements on the hysteretic behavior
hinged to the first-order CDW transition have been already accessible
in current experiments \citep{Landig_Nature_2016}, identifying the
existence of the critical point experimentally can thus be greatly
facilitated via monitoring the disappearance of the hysteretic behavior
upon increasing temperature from the low-temperature regime. We believe
our work will stimulate further experimental and also theoretical
investigations on possible emergent criticalities under the influence
of both thermal fluctuations and ILR interactions, especially beyond
the deep Mott-insulator limit. 
\begin{acknowledgments}
This work was supported by NSFC (Grant No.~11874017, No.~11674334,
and No.~11947302), GDSTC under Grant No.~2018A030313853, Science
and Technology Program of Guangzhou (Grant No.~2019050001), and START
grant of South China Normal University. 
\end{acknowledgments}

\appendix

\section{Hubbard-Stratonovich transformation on the partition function\label{sec:Appendix_HST}}

To investigate finite temperature properties of the system, the central
quantity we need to calculate is the quantum partition function $Z=\mathrm{tr}\exp[-\beta(\hat{H}-\mu\hat{N})]$
with $\hat{N}=\sum_{i,\sigma}\hat{n}_{i,\sigma}$ and $\mu$ being
the chemical potential. Its explicit form in the occupation number
representation reads\begin{widetext} 
\begin{align}
Z & =\sum_{\{n_{i,\sigma}\}}e^{-\beta\left\{ \sum_{i,\sigma}\left[\frac{U_{s}}{2}n_{i,\sigma}(n_{i,\sigma}-1)-\mu n_{i,\sigma}\right]-\frac{U_{l}}{L}\left[\sum_{i}\left(n_{i,e}-n_{i,o}\right)\right]^{2}\right\} },\label{eq:Partition_function_explicit_form}
\end{align}
where $n_{i,\sigma}$ is the occupation number, i.e., the eigenvalue
of the bosonic particle number operator $\hat{n}_{i,\sigma}$.

The Hubbard-Stratonovich transformation that we use to introduce the
CDW order parameter field $\phi$ into the partition function reads
\begin{align}
\left(\sqrt{\frac{\beta U_{l}L}{\pi}}\right)^{-1}\exp\left(\beta\frac{U_{l}}{L}\left[\sum_{i=1}^{L/2}\left(n_{i,e}-n_{i,o}\right)\right]^{2}\right) & =\int_{-\infty}^{+\infty}d\phi\,\exp\left(-\beta U_{l}\left\{ L\phi^{2}+2\left[\sum_{i=1}^{L/2}\left(n_{i,e}-n_{i,o}\right)\right]\phi\right\} \right).\label{eq:HST}
\end{align}
By using Eq.~(\ref{eq:HST}) we can replace the long range interaction
term appearing in Eq.~(\ref{eq:Partition_function_explicit_form})
by the integral over the $\phi$ field and rewrite the partition function
$Z$ in terms of $\phi$ as shown in Eq.~(2)\textcolor{blue}{{}
}in the main text.

$\phi$ assumes the physical meaning of the fluctuating CDW order
parameter filed. It appears in the partition function and its expectation
value $\bar{\phi}\equiv\langle\phi\rangle$ equals to CDW order parameter
$\langle\sum_{i=1}^{L/2}\hat{n}_{i,e}-\sum_{i=1}^{L/2}\hat{n}_{i,o}\rangle/L$
exactly. This can be shown by introducing a source $J$ that couples
to $L^{-1}\sum_{i=1}^{L/2}(n_{i,e}-n_{i,o})$ in the partition function.
Now the partition function depends on the source $J$ and its explicit
form reads

\begin{align}
Z[J]= & \sqrt{\frac{\beta U_{l}L}{\pi}}\int_{-\infty}^{+\infty}d\phi\,\sum_{\{n_{i,\sigma}\}}e^{-\beta\sum_{i,\sigma}\left[\frac{U_{s}}{2}n_{i,\sigma}(n_{i,\sigma}-1)-\mu n_{i,\sigma}\right]}e^{-\beta U_{l}L\phi^{2}+2\beta U_{l}\phi\sum_{i=1}^{L/2}\left(n_{i,e}-n_{i,o}\right)}e^{JL^{-1}\sum_{i=1}^{L/2}(n_{i,e}-n_{i,o})}
\end{align}
One can directly show 
\begin{equation}
\frac{\langle\sum_{i=1}^{L/2}\hat{n}_{i,e}-\sum_{i=1}^{L/2}\hat{n}_{i,o}\rangle}{L}=\left.\frac{\partial\ln Z[J]}{\partial J}\right|_{J=0}.\label{eq:derivative_wrt_source_before_shift}
\end{equation}
Moreover, since $\phi$ is an integral variable with its domain lies
in $(-\infty,+\infty)$, one can shift $\phi$ without change the
partition function. After the shift $\phi\rightarrow\phi-\frac{J}{2\beta U_{l}L}$,
we get 
\begin{align}
Z[J] & =\sqrt{\frac{\beta U_{l}L}{\pi}}\int_{-\infty}^{+\infty}d\phi\,\sum_{\{n_{i}\}}e^{-\beta\sum_{i,\sigma}\left[\frac{U_{s}}{2}n_{i,\sigma}(n_{i,\sigma}-1)-\mu n_{i,\sigma}\right]}e^{-\beta U_{l}L\phi^{2}+2\beta U_{l}\phi\sum_{i=1}^{L/2}(n_{i,e}-n_{i,o})+\phi J-\frac{J^{2}}{4\beta U_{l}L}}.
\end{align}

\end{widetext} Now one can calculate the same derivative $\left.\partial\ln Z[J]/\partial J\right|_{J=0}$
and get 
\begin{equation}
\left.\frac{\partial\ln Z[J]}{\partial J}\right|_{J=0}=\langle\phi\rangle,\label{eq:derivative_wrt_source_after_shift}
\end{equation}
Comparing Eq.~(\ref{eq:derivative_wrt_source_before_shift}) to Eq.~(\ref{eq:derivative_wrt_source_after_shift}),
we conclude $\phi$ assumes the physical meaning of fluctuating CDW
order parameter field, with 
\begin{equation}
\langle\phi\rangle=L^{-1}\langle\sum_{i=1}^{L/2}\hat{n}_{i,e}-\sum_{i=1}^{L/2}\hat{n}_{i,o}\rangle.
\end{equation}

\section{Ginzburg-Landau effective theory\label{sec:GL-theory}}

In the critical regime, the CDW order parameter is small enough to
allow a systematic expansion of the system's free energy with respect
to its CDW order parameter $\bar{\phi}$. The Ginzburg-Landau free
energy $F$ that is allowed by the $\mathbb{Z}_{2}$ symmetry (i.e.,
$F$ should be invariant under the transformation $\bar{\phi}\rightarrow-\bar{\phi}$)
assumes the form 
\begin{equation}
F=\frac{1}{2}r\bar{\phi}^{2}+\frac{1}{4}u_{4}\bar{\phi}^{4}+\frac{1}{6}u_{6}\bar{\phi}^{6},
\end{equation}
where we expand $F$ up to the sixth order in $\bar{\phi}$, with
$r$, $u_{4}$, and $u_{6}$ being the GL coefficients.

\subsection{Critical scaling of the CDW order parameter jump $\Delta\bar{\phi}$
in the vicinity of the critical point}

To describe the first-order transition, one further assume: $r$ depends
on temperature linearly, i.e., $r=a(T_{\mathrm{CP}}-T)$ where $a$
is a positive coefficient, and $u_{4,6}$ are temperature independent
coefficients with $u_{4}<0$ for $T<T_{\mathrm{CP}}$ and $u_{6}$
being always positive in order to stabilize the whole system. With
$u_{4}<0$, the GL free energy assumes three minimums located at 
\begin{equation}
\bar{\phi}=0,\pm\sqrt{\frac{-u_{4}+\sqrt{u_{4}^{2}-4u_{6}r}}{2u_{6}}},
\end{equation}
respectively. For the system parameters located at the first-order
transition boundary, the conditions 
\begin{equation}
F=0\mathrm{\,\,and\,\,}\frac{\partial F}{\partial\bar{\phi}}=0,
\end{equation}
should both hold true, from which we can obtain the non-zero CDW order
assumes the values 
\begin{equation}
\bar{\phi}=\pm\frac{1}{2}\sqrt{\frac{r}{|u_{4}|}}.
\end{equation}
This thus indicates $\Delta\bar{\phi}=(1/2)\sqrt{r/\left|u_{4}\right|}$.
Noticing $r=a(T_{\mathrm{CP}}-T)$ with $T$ being the transition
temperature at the first-order transition boundary, this gives rise
to the power law scaling 
\begin{equation}
\Delta\bar{\phi}\propto(T_{\mathrm{CP}}-T)^{1/2}.
\end{equation}

\subsection{Critical scaling of the CDW order parameter $\bar{\phi}$ of the
second-order CDW transition}

Above the critical point, the CDW transition is a second-order phase
transition, thus $u_{4}$ is assumed to be positive. In this case,
the sixth order term in GL free energy is irrelevant in the vicinity
of the transition, therefore we only need to keep up to the fourth
order term in the GL free energy, i.e., 
\begin{equation}
F=\frac{1}{2}r\bar{\phi}^{2}+\frac{1}{4}u_{4}\bar{\phi}^{4}.
\end{equation}
As the first-order transition case, we assume $r=a(T-T_{c})$ with
$T_{c}$ being the critical temperature of the second-order CDW transition.
The CDW order is determined by the condition 
\begin{equation}
\frac{\partial F}{\partial\bar{\phi}}=0,
\end{equation}
from which we obtain 
\begin{equation}
\bar{\phi}=\sqrt{\frac{|r|}{u_{4}}},
\end{equation}
for $r<0$. Indeed, noticing $r=a(T-T_{c})$, this gives rise to the
power scaling 
\begin{equation}
\bar{\phi}\propto(T_{c}-T)^{1/2}.
\end{equation}

\end{document}